\documentclass[twocolumn,floatfix,superscriptaddress,a4paper,showkeys,nofootinbib,reprint]{revtex4-1}
\usepackage[colorlinks=true,linktocpage=true,linkcolor=blue,citecolor=blue,allcolors=blue]{hyperref}
\usepackage{epsfig}
\usepackage{latexsym}
\usepackage[utf8]{inputenc}
\usepackage{xspace}
\usepackage{indentfirst}
\usepackage{enumitem}
\usepackage{color}
\usepackage{placeins}
\usepackage[table,xcdraw]{xcolor}
\usepackage{setspace}
\usepackage{lipsum}
\usepackage{multirow}
\usepackage{hyperref}

\usepackage{graphicx}

\usepackage{todonotes}

\usepackage{amsmath}
\usepackage{amssymb}
\usepackage[english]{babel}
\usepackage{url}
\graphicspath{{figs/}}
\newcommand{\mean}[1]{\langle #1 \rangle}
\newcommand{\eq}[1]{\begin{align} #1 \end{align}}

\newcommand{\be}{\begin{equation}}
\newcommand{\ee}{\end{equation}}

\begin{document}

\title{Molecular dynamics analysis of particle number fluctuations in the mixed phase of a first-order phase transition}
\author{Volodymyr~A.~Kuznietsov}
    \affiliation{Bogolyubov Institute for Theoretical Physics, 03680 Kyiv, Ukraine}   
    \affiliation{Leibniz Institute For Solid State and Material Research, 01069 Dresden, Germany,}
    \affiliation{Frankfurt Institute for Advanced Studies, Giersch Science Center, Ruth-Moufang-Str. 1, D-60438 Frankfurt am Main, Germany}
\author{Oleh Savchuk}
\affiliation{Facility for Rare Isotope Beams, Michigan State University, East Lansing, MI 48824 USA}
\affiliation{Bogolyubov Institute for Theoretical Physics, 03680 Kyiv, Ukraine}
\affiliation{
GSI Helmholtzzentrum f\"ur Schwerionenforschung GmbH, Planckstr. 1, D-64291 Darmstadt, Germany}
   \affiliation{Frankfurt Institute for Advanced Studies, Giersch Science Center, Ruth-Moufang-Str. 1, D-60438 Frankfurt am Main, Germany}
   
\author{Roman~V.~Poberezhnyuk}
   \affiliation{Bogolyubov Institute for Theoretical Physics, 03680 Kyiv, Ukraine}
   \affiliation{Frankfurt Institute for Advanced Studies, Giersch Science Center, Ruth-Moufang-Str. 1, D-60438 Frankfurt am Main, Germany} 

\author{Volodymyr~Vovchenko}
\affiliation{Physics Department, University of Houston, Box 351550, Houston, TX 77204, USA}
\affiliation{Nuclear Science Division, Lawrence Berkeley National Laboratory, 1 Cyclotron Road, Berkeley, California 94720, USA}

\author{Mark~I.~Gorenstein}
    \affiliation{Bogolyubov Institute for Theoretical Physics, 03680 Kyiv, Ukraine}
    \affiliation{Frankfurt Institute for Advanced Studies, Giersch Science Center, Ruth-Moufang-Str. 1, D-60438 Frankfurt am Main, Germany}

\author{Horst~Stoecker}
   \affiliation{Frankfurt Institute for Advanced Studies, Giersch Science Center, Ruth-Moufang-Str. 1, D-60438 Frankfurt am Main, Germany}
    \affiliation{
GSI Helmholtzzentrum f\"ur Schwerionenforschung GmbH, Planckstr. 1, D-64291 Darmstadt, Germany}
    \affiliation{Institute of Theoretical Physics, Goethe Universit\"at, Frankfurt, Germany}  
   
\date{\today}

\begin{abstract}

Molecular dynamics simulations are performed for a finite
non-relativistic system of particles with Lennard-Jones potential. We study the effect of liquid-gas mixed phase on particle number fluctuations in coordinate subspace. 
A metastable region of the mixed phase, the so-called nucleation region, is analyzed in terms of a non-interacting cluster model. 
Large fluctuations due to spinodal decomposition are observed. 
They arise due to the interplay between the size of the acceptance region and that of the liquid phase.
These effects are studied with a simple geometric model. 
The model results for the scaled variance of particle number distribution are compared with those obtained from the direct molecular dynamic simulations. 

\end{abstract}
\maketitle

\section{Introduction}
The endpoint of a first-order phase transition, noted as the critical point (CP), occurs under different physical conditions, including most molecular and ferromagnetic systems~\cite{LandauBook,GreinerBook}, nuclear matter \cite{allen2017computer}, and potentially the hot QCD matter at nonzero baryon density \cite{Stephanov:1999zu,Bzdak:2019pkr}. 
In the thermodynamic limit, particle number fluctuations exhibit singular behavior at the CP.
These singularities are smeared out in finite-size systems.
Nevertheless, small systems also demonstrate  
specific features of critical behavior such as enhancement of fluctuations \cite{Bernhardt2022,Kuznietsov2022}.

Event-by-event fluctuations in nucleus-nucleus collisions are used as an experimental tool to search for the QCD CP at finite baryon density \cite{Stephanov:1999zu,Bzdak:2019pkr}.
The presence of the QCD CP should manifest itself in the enhanced fluctuations of proton number~\cite{Hatta:2003wn} and possibly non-monotonic collision energy dependence of non-Gaussian fluctuation measures~\cite{Stephanov:2008qz,Stephanov:2011pb}.
Measurements of proton number fluctuations in nucleus-nucleus collisions have been performed by different experiments such as STAR~\cite{STAR:2020tga,STAR:2021iop}, HADES~\cite{HADES:2020wpc}, and ALICE~\cite{ALICE:2019nbs}.
The measurements indicate a possible non-monotonic collision energy dependence of the kurtosis of proton number~\cite{STAR:2020tga} as well as a possible enhancement of two-proton correlations over non-critical baselines~\cite{Vovchenko:2021kxx} but conclusive evidence for the presence of QCD CP is still lacking.

\begin{figure}[t]
	\includegraphics[scale=0.55]{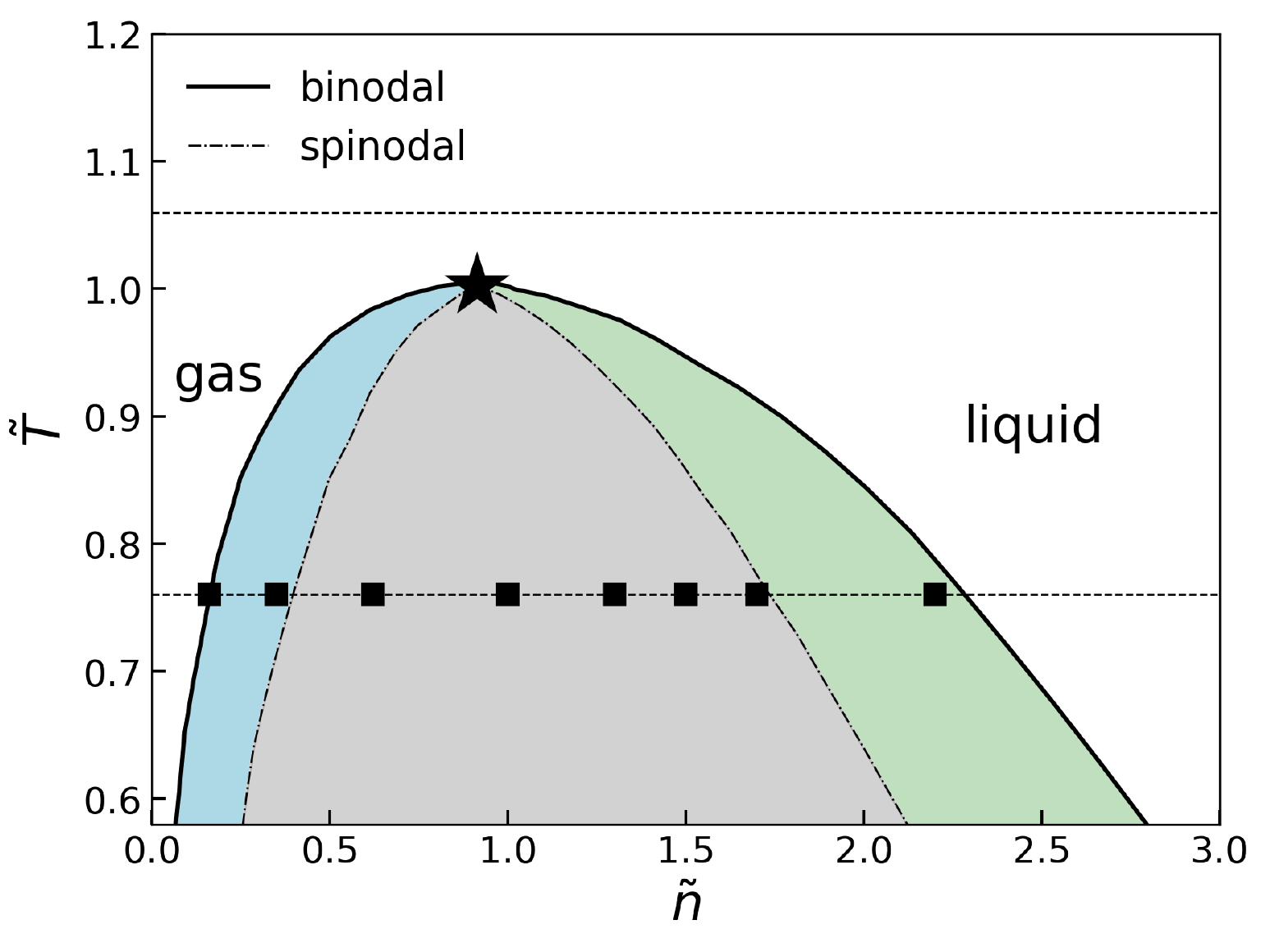}
\caption{The liquid-gas region of the Lennard-Jones fluid phase diagram. 
 Horizontal dashed lines show the subcritical isotherm $\tilde T = 0.76$ studied in this work and the supercritical isotherm $\tilde T = 1.06$ explored in Ref.~\cite{Kuznietsov2022}.
 Solid and dashed lines show the binodal and spinodal lines, respectively.
 The blue and green regions correspond to the nucleation and cavitation metastable parts of the mixed phase, respectively.
 The grey area denotes the spinodal decomposition region.
 The black star represents the CP.
 The squares denote the $(\tilde{n}, \tilde{T})$ points where 
 the MD simulations in the mixed phase have been performed.
 }
    \label{fig:pd}
\end{figure}

The
grand canonical ensemble (GCE) of statistical mechanics is the most suitable framework to study statistical fluctuations. Within this formulation, the cumulants of particle number distribution are straightforwardly connected to the chemical potential derivatives of thermodynamic potential. 
However, the GCE can not be directly used for the conditions realized in the experiment~\cite{Koch:2008ia,Vovchenko:2021gas}. 
Several essential restrictions should be taken into account: 
(i) finite size
of systems created in the experiment~\cite{Berdnikov:1999ph,Poberezhnyuk:2020ayn}, 
(ii) influence of the  global conservation laws, for instance, baryon number conservation~\cite{Bzdak:2012an,Vovchenko:2020tsr}, and 
(iii) differences between coordinate and momentum space acceptances.
Recently the subensemble acceptance method~(SAM) to correct the fluctuation measurements for global conservation laws has been  developed~\cite{Vovchenko:2020tsr,Vovchenko:2020gne,Poberezhnyuk:2020ayn,Vovchenko:2021yen,Barej:2022jij,Barej:2022ccb}. 
This method is applicable for statistical  systems in the presence of interactions.
In the limit of ideal Maxwell-Boltzmann gas, it reduces to the binomial acceptance correction procedure \cite{Bzdak:2012an,Braun-Munzinger:2016yjz,Savchuk2019xfg}.

In the present work we continue our studies \cite{Kuznietsov2022} of particle number fluctuations within molecular dynamics~(MD) simulations of the Lennard-Jones~(LJ) fluid. 
The model considered here corresponds to an interacting system of non-relativistic particles.  
The presence of both attractive and repulsive interactions leads to a first-order liquid-gas phase transition (LGPT). 
The MD simulations of the LJ fluid provide a microscopic approach to fluctuations in a system with a phase transition. They also allow one to study deviations from the baselines based on the GCE. 
This study thus complements earlier analyses of correlations and fluctuations in the first-order phase transition region performed using hadronic transport with mean fields~\cite{Sorensen:2020ygf,Savchuk:2022msa} or fluid dynamics with a finite-range term~\cite{Steinheimer:2012gc,Steinheimer:2013gla}.
With regard to mean quantities the molecular dynamics of non-equilibrium finite systems was studied previously in the context of heavy ion collisions in Refs.~\cite{Molitoris:1984xv,Peilert:1987xb,Horst1988_qm,PhysRevC.39.1402,Hartnack:1997ez}.

Our study is motivated by the measurements of baryon number fluctuations in heavy-ion collisions to probe the QCD phase structure.
In particular, the LJ fluid can naturally model the nuclear liquid-gas transition between a dilute gas of nucleons and clusters and the dense nuclear liquid, if one regards the LJ particles as nucleon degrees of freedom.
This nuclear LGPT is probed in nuclear collisions at low energies~\cite{Pochodzalla:1995xy,Natowitz:2002nw,Karnaukhov:2003vp}.
Experiments at higher collision energies, on the other hand, study the confinement-deconfinement transition, which may contain a critical point and a line of first-order phase transition at finite baryon density~\cite{Stephanov:1999zu,Bzdak:2019pkr}.
The relevance of the LJ fluid to model the confinement-deconfinement transition may seem less evident, given that it does not describe the expected change of degrees of freedom from hadrons to quarks.
Nevertheless, simulations of the LJ fluid do provide useful guidance to understand the behavior of baryon number fluctuations near the QCD CP, for two reasons: (i) the behavior of baryon number fluctuations is universal near the QCD CP and governed by the 3D-Ising universality class~\cite{Bzdak:2019pkr} -- the same universality class that characterizes critical behavior in the LJ fluid~\cite{caillol1998critical}; (ii) the LJ fluid simulations can test the validity of the model-independent SAM procedure for subtracting the canonical ensemble effects on baryon number cumulants, this is particularly relevant given that the finite-size effects, that hinder the accuracy of the SAM, can be significant in the mixed phase region.

This work focuses on fluctuations in the mixed-phase region of a first-order phase transition.
While significant attention has been given to higher-order measures of fluctuations of conserved charges at supercritical temperatures and in pure phases~(see e.g. Refs.~\cite{Vovchenko:2015pya,Stephanov:1999zu,Hatta:2002sj,Stephanov:2008qz,Stephanov:2011pb,Mukherjee:2016nhb,Poberezhnyuk:2019pxs,Motornenko:2019arp}), less attention has been paid to the mixed phase. However, it is possible for a system created in relativistic nucleus-nucleus collisions to enter the mixed phase of a first-order phase transition under certain conditions. This is particularly relevant because of the ongoing program of the HADES collaboration  at the GSI Helmholtzzentrum f{\"u}r Schwerionenforschung mbH to measure higher-order net-proton and net-charge fluctuations in central Au+Au collisions at collision energies of $0.2A -1.0A$~GeV. The system created in these collisions may undergo freeze-out in the mixed phase of the nuclear LGPT.


In our previous work~\cite{Kuznietsov2022}, we studied a supercritical isotherm, $T=1.06\,T_c$, observing a sizable increase of particle number fluctuations near the critical particle number density $n\approx n_c$. 
In the present work, we study particle number fluctuations along a subcritical temperature $T=0.76\,T_c$ inside the liquid-gas mixed phase. 
First, we look at the metastable part of the mixed phase -- the so-called nucleation region.   
The simulation results are compared to a simple model of non-interacting particle clusters.  
Another part of the liquid-gas mixed phase -- the spinodal decomposition region -- demonstrates anomalous large particle number fluctuations. This happens at temperature $T$ and particle number density $n$ also far away from the CP.
A simple analytical toy model is constructed to clarify these effects.
\begin{figure*}[t]
    \includegraphics[scale=0.45]{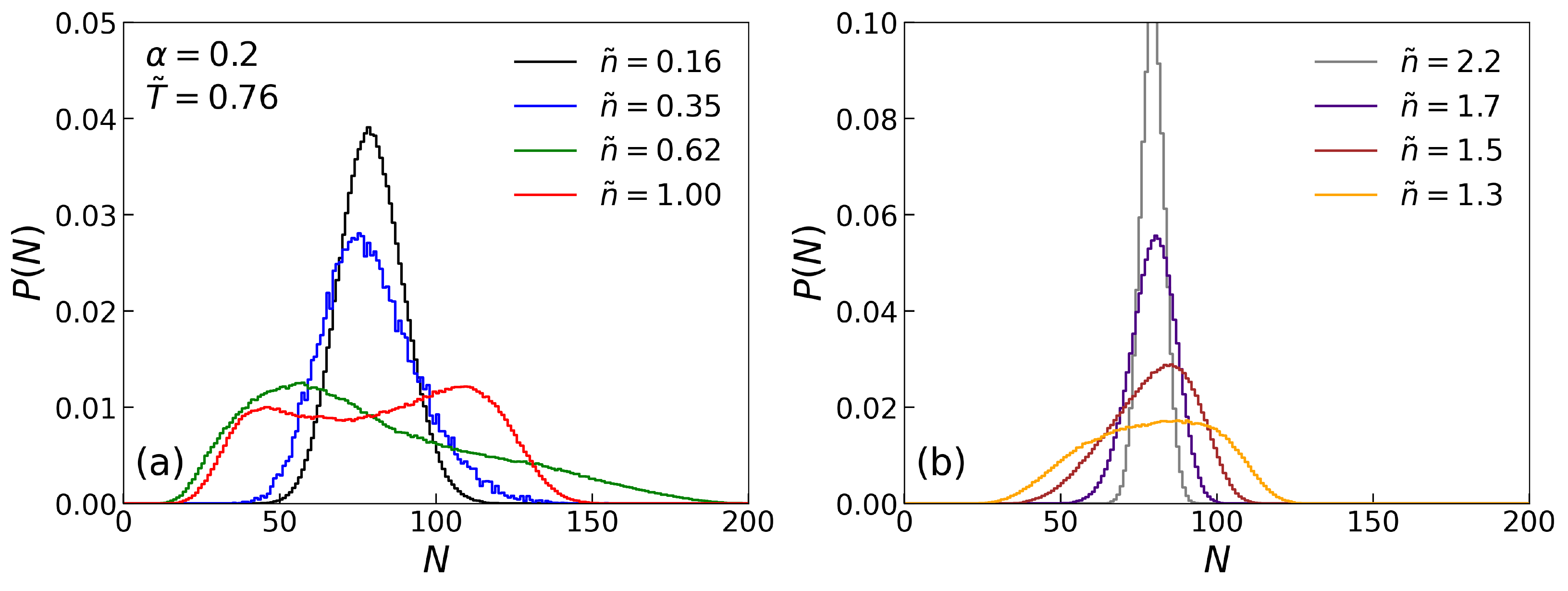}
    \caption{The particle number distributions in a subvolume $V=\alpha V_0$ for the system with $N_0=400$ particles. The distributions $P(N)$ obtained from the MD simulations at $\tilde T=0.76$ and different values of $\tilde n$ inside the mixed phase. }
    \label{fig:p1}
\end{figure*}

The paper is organised as follows. The details of MD  
with LJ potential and  the results of the simulations for particle number fluctuations are presented in Sec. \ref{MDLJ}. A brief description of the mixed phase structure is described in Sec. \ref{mixp}. A simple model of non-interacting clusters  in Sec.~\ref{CM} and a geometrical toy model in Sec.~\ref{SR} are developed to interpret the MD results in the nucleation and spinodal decomposition regions,
respectively.
Summary in Sec. \ref{res} closes the article

\section{Molecular dynamics with Lennard-Jones potential}\label{MDLJ}
We use molecular dynamics simulations of the classical non-relativistic   system of particles interacting via the Lenard-Jones (LJ) potential,
\eq{V_{\rm LJ}(r) = 4\varepsilon\left[\left(\frac{\sigma}{r}\right)^{12} - \left(\frac{\sigma}{r}\right)^{6}\right]~. \label{LJ}
}
The first term in Eq.~(\ref{LJ}) corresponds to the repulsive forces at short distances whereas the second one describes the attractive interactions.
The  parameter $\epsilon$ describes the depth of the attractive well, and $\sigma$ corresponds to the size of the particle, which also defines the distance scale. 
It is convenient to introduce dimensionless
reduced variables, 
\eq{
V^*_{\rm LJ}(r^*) = V_{\rm LJ}(r)/\varepsilon = 4\left((r^*)^{-12} -  (r^*)^{-6} \right)~,
}
with $r^* = r/\sigma$ being the reduced distance.
The reduced thermodynamic variables are the temperature $T^* = T/(\varepsilon)$, particle number density $n^* = n \sigma^3$, and pressure $p^*=p\sigma^3/\varepsilon$. 
The particle's mass can be utilized to define the dimensionless time variable, $t^* = t \sqrt{\varepsilon /(m\sigma^2)}$.

\begin{figure*}[t]
	\includegraphics[scale=0.45]{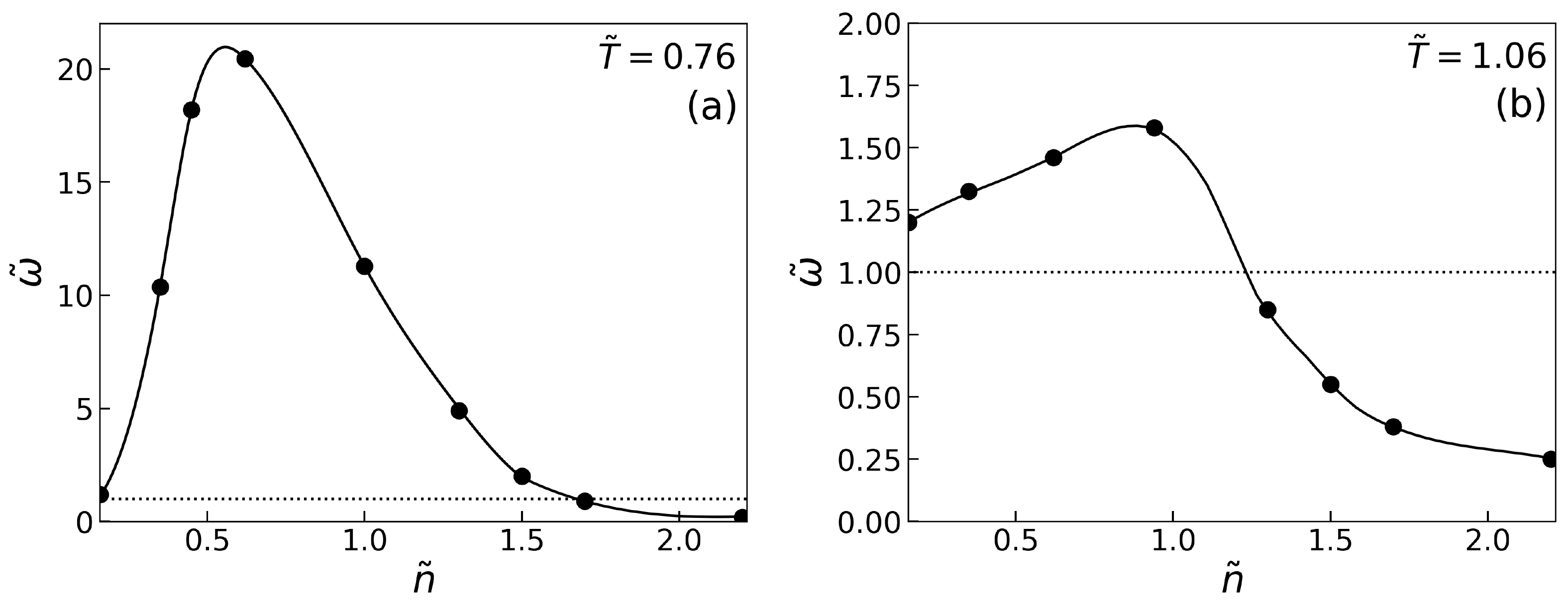}
	\caption{The scaled variance $\tilde \omega$ (\ref{wgce}) 
 as a function of the particle number density $\tilde n$ for $N_0=400$, $\alpha =0.2$ at $\tilde T=0.76$ (a) and $\tilde T= 1.06$ (b). 
 }
    \label{fig:w0}
\end{figure*}

The LJ  system possesses a rich phase diagram
(see e.g. Ref.~\cite{STEPHAN2020112772} for an overview).
At present, there are no direct analytical tools to compute the phase diagram in the LJ system. Nevertheless, numerical methods (see, e.g. Ref.~\cite{Evans1979}) allow one to compute the approximate locations of the LGPT binodal and spinodal lines, as well as the CP location. This part of the phase diagram is of primary interest in the present work, and it is shown in Fig.~\ref{fig:pd} in terms of the reduced temperature and density.
The CP location has been estimated from numerous
MD simulations 
\cite{doi:10.1021/acs.jcim.9b00620}
\eq{
\label{eq:TcLJ} 
\begin{split}
T^*_c &  = 1.321 \pm 0.007~ , ~~\\
n^*_c & = 0.316 \pm 0.005~, ~~\\
p^*_c & = 0.129 \pm 0.005~.
\end{split}
}
In what follows, we use a set of dimensionless variables scaled by the critical values
\eq{\tilde T~\equiv \frac{T}{T_c} = \frac{T^*}{T^*_{\rm c}}~,~~~\tilde n \equiv \frac{n}{n_c}= \frac{n^*}{n^*_{\rm c}}~,~~~\tilde p \equiv \frac{p}{p_c}=\frac{p^*}{p^*_{\rm c}}~. \label{T-tilde}
} 
The quantities (\ref{eq:TcLJ}) correspond to the thermodynamic limit when
the system's volume $V\rightarrow\infty$. 
For finite systems, the physical meaning of the LGPT and its CP should be treated with caution, 
as they are only rigorously defined for infinite systems.

The MD simulations are performed by numerically integrating Newton's equations of motion using the Velocity Verlet integration method.
The simulations are done for a system of $N_0=400$ interacting particles in a cubic box of volume $V_0$ with periodic boundary conditions with minimum image convention.

In the mixed phase the time of reaching the thermal equilibrium can be rather large (see Refs. \cite{Sanz_2013}). 
After the equilibration time, $\tilde t_{\rm eq} = 50$, the LJ system reaches a state with a stable temperature\footnote{During all system evolution some temperature fluctuations can be seen, but they are relatively small, so the mean value of temperature differs from the desired by no more than $0.4\%$.} (see Ref.~\cite{Kuznietsov2022}).
The time of all simulations is $\tau = 10^6$.
This large time interval guarantees small deviations (less than 1\%) of the scaled variance in independent simulations.

The total particle number $N_0$ in the entire volume is fixed. 
To study the fluctuations of particle number one thus needs
to choose a subvolume $V=\alpha V_0$
 ($0<\alpha<1$) of the whole volume.
We choose a cubic subvolume 
placed in the geometrical center of the system.
From the MD simulations, we obtain the normalized probability distribution $P(N)$ to observe $N$ particles in the subvolume $V$.

\begin{figure*}[t]
    \includegraphics[width=1\textwidth]{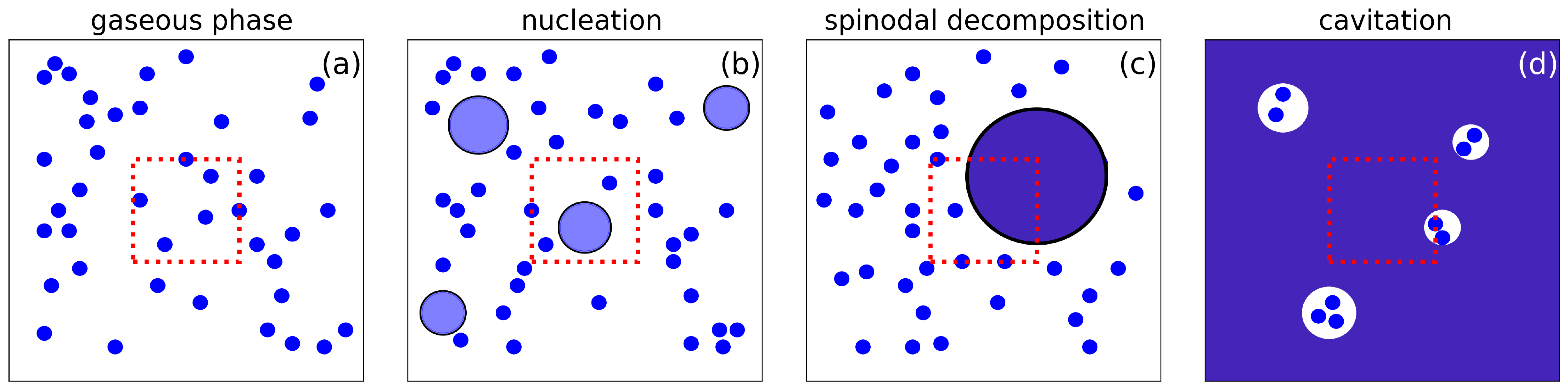}
    \caption{
    Different regions along the supercritical isotherm of the liquid-gas transition:
    (a) gaseous phase, (b) nucleation, (c) spinodal decomposition, and (d) cavitation.
    }
    \label{fig:mixed}
\end{figure*}

\begin{figure*}[t]
    \includegraphics[width=0.49\textwidth]{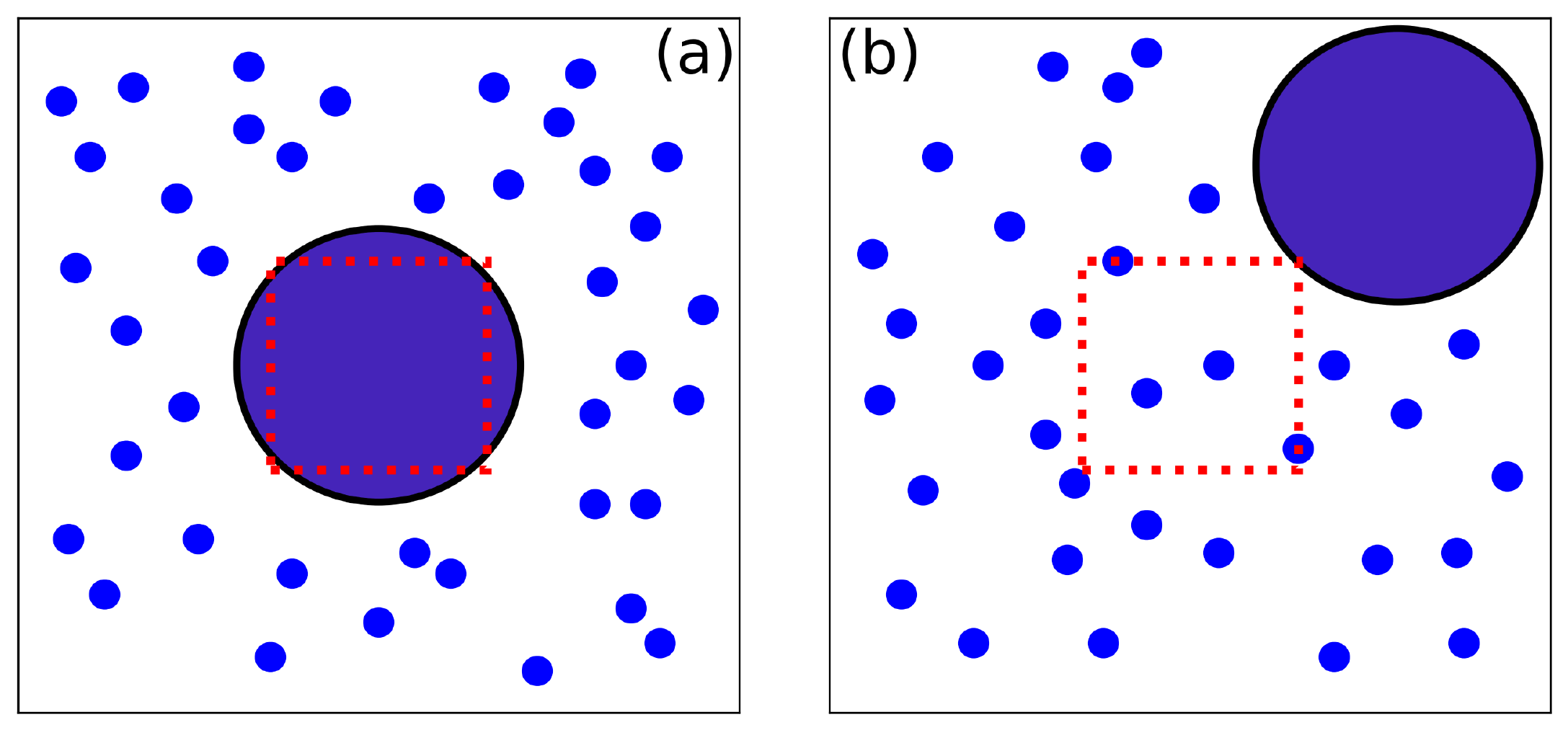}
    \caption{
    Possible position of the liquid phase in the spinodal decomposition region relative to the acceptance subvolume (red dashed square).
    }
    \label{fig:blob}
\end{figure*}

A useful measure of particle number fluctuations is the scaled variance:
\eq{\omega = \frac{\left<N^2\right> - \left<N\right>^2}{\left<N\right>}.\label{omega}}
In MD simulations, the values $\left<N\right>$ and $\left<N^2\right>$ can be calculated as time averages.
In Fig.~\ref{fig:p1} we present the $P(N)$ distribution 
at the subcritical temperature $\tilde T=0.76$ for several different particle number densities $\tilde n$ inside the mixed phase. 
The total number of particles is fixed as $N_0=400$ and the subvolume fraction is taken as $\alpha = 0.2$.
From Fig. \ref{fig:p1}, one observes substantial deviations of the resulting distributions from
the Poisson distribution baseline.
For $\tilde n\approx 1$, a double-hump distribution is clearly observed.

Note that for any finite $\alpha$, fluctuations of $N$ in the subvolume $V$ will be influenced by the exact global conservation of the total particle number $N_0$ in the full volume $V_0$.  
In the large volume limit, these effects can be taken into account analytically~\cite{Vovchenko:2020tsr}.
One can defined a scaled variance $\tilde{\omega}$ corrected for exact $N_0$ conservation as
\eq{\tilde \omega = \frac{\omega}{1 - \alpha}~.\label{wgce}}

The results for the corrected scaled variance $\tilde \omega$ as a function of $\tilde n$ are presented in Fig.~\ref{fig:w0} for both (a) the subcritical and (b) the supercritical isotherms $\tilde T=0.76$ and $\tilde T = 1.06$, respectively.
All results are obtained for $N_0=400$ and $\alpha=0.2$, as in Fig.~\ref{fig:p1}.

One can immediately observe that fluctuations are much larger in the mixed phase at $\tilde T= 0.76$ compared to those along the  temperature $\tilde T=1.06$, slightly above the critical point.
This indicates that, although the fluctuations exhibit singular behavior in the vicinity of the CP, they can be even larger in the mixed phase region away from the critical point.

In the following sections, we provide a brief overlook of the structure of the liquid-gas mixed phase and 
analyze the observed large values of $\tilde\omega$ in the mixed phase in terms of simple analytical models.

\section{Mixed phase structure}
\label{mixp}

One can specify three different regions inside the mixed phase:
nucleation, spinodal decomposition, and cavitation (see, e.g., Refs.~\cite{frenkel1946,Fisher1967}). 
They are shown in Fig.~\ref{fig:pd} by blue, grey, and green colors, respectively.  
Their microscopic structures are symbolically illustrated in Fig.~\ref{fig:mixed}. The nucleation region includes a mixture of particles and small clusters (liquid droplets), whereas the cavitation region 
is represented by the liquid with small bubbles of the gaseous phase.
In the context of heavy ion collision clusters correspond to nuclear fragments whose distributions were previously studied using MD in the case of expanding system in Refs.~\cite{Molitoris:1984xv,STOCKER1986277,Peilert:1987xb,Horst1988_qm,PhysRevC.39.1402}. Experimental measurements of nuclear fragment mass distributions were used to probe the nuclear LGPT and the CP (see, e.g., Refs.~\cite{Lindenstruth:1993ax,lindenstruth1993dynamical,Pochodzalla:1995xy,Pochodzalla:1997vq}).
The nucleation and cavitation regions of the mixed phase correspond to the metastable states. 
In the MD simulations one expects to achieve an equilibrated {\it steady state} in these regions after a sufficiently long time.
In most cases, however, the time to reach complete equilibrium appears very long.  
Note also that a strict physical meaning and location of the bounds of different regions are  dependent on the size of the system (see, e.g.,  Refs. \cite{Klein1983,Csernai:1994nh,Wedekind2009}) and can be sensitive to the collective motion \cite{Pratt1990,Kunde:1994sw,Pratt:1995cqc}. 

The spinodal decomposition region is fundamentally different from the metastable nucleation and cavitation ones (see, e.g., Refs.~\cite{Lopez1989,Favvas2008}).
The LGPT manifests itself here as a fast system separation into the  gaseous and liquid phases. 
The equilibrium states in this region (see, e.g., Ref. \cite{Elliott1989}) are achievable in the MD simulations.
The heterogeneous structure of the spinodal decomposition phase is illustrated in Fig.~\ref{fig:blob}, showing a strong influence on the particle number fluctuations obtained in the MD simulations. This is discussed in more detail in Sec.~\ref{SR}. 
One can note a principal difference between the heterogeneous two-phase states in the spinodal decomposition region and the homogeneous mixtures of particles plus clusters in the nucleation region
and liquid with gaseous bubbles in the cavitation region. 

\section{Mixture of particles and Clusters}\label{CM}

\begin{figure*}[t]
    \includegraphics[scale=0.5]{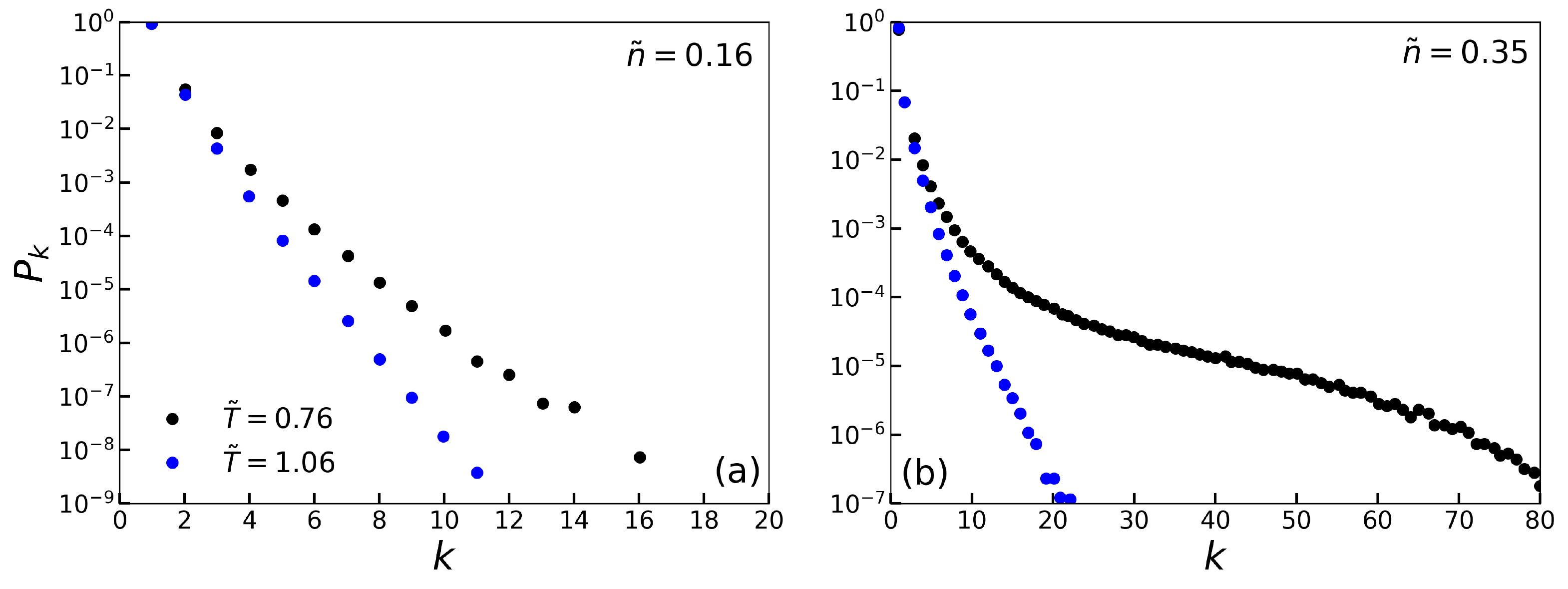}
    \caption{Cluster probability distributions $P_k$
    extracted from the MD of Lennard-Jones fluid for $N_0=400$
    and $\tilde T=0.76$ at gaseous binodal $\tilde n=0.16$
    (a) and gaseous spinodal $\tilde n=0.35$ (b). For comparison, $P_k$ distributions are also presented for supercritical temperature $\tilde T=1.06$. 
    }
    \label{fig:Pm1}
\end{figure*}

To clarify some general features of the nucleation region, let us consider   
a non-interacting multi-component gas of $k$-particle clusters~($k=1,2,\ldots$).
The GCE
partition function reads
\eq{\begin{split}
&Z_{\rm GCE} =
\prod_{k \geq 1}  \sum_{N_k = 0}^{\infty} \frac{\left(Vg(k)e^{\mu k/T}\right)^{N_k}}{N_k!}(2\pi k m T)^{3N_k/2} \\&= \prod_{k\ge 1}  \exp\left[V(2\pi k m\, T)^{3/2}\, g(k)\,\exp\left(\frac{\mu k 
}{T}\right) \right],
\end{split}\label{Zgce0}}
where $V$, $T$, and $\mu$ are, respectively, the system volume, temperature, and chemical potential that corresponds to the total conserved number $N$ of particles over all clusters; $g(k)$ is the ``degeneracy"  factor (number of internal states of the $k$-th cluster), and $m$ the mass of a single particle, such that the mass of a $k$-particle cluster equals $M_k = k m$). 
The system is considered to be in chemical equilibrium, thus $\mu_k = k \mu$.
The CE partition function $Z_{\rm CE}(V,T,N)$ of the cluster model (\ref{Zgce0})
is considered in
Appendix~\ref{A}, where it is shown that the moments $\mean{k^l}$ of the cluster distribution are identical between the CE and the GCE in the thermodynamic limit.

The cluster distribution (i.e., the normalized probability to find  the $k$th cluster in the cluster system)  can be written in a form

\eq{P_k(T,\mu) \equiv \frac{\left<N_k\right>}{\sum\limits_{l \geq 1} \left< N_l\right>} = \frac{k^{3/2}g(k)\exp\left(\frac{\mu k}{T}\right)}{\sum\limits_{l \geq 1} l^{3/2}g(l)\exp\left(\frac{\mu l}{T}\right)}~, \label{Pk}
}
where 
\eq{k\langle N_k \rangle = \frac{\partial \ln \left[Z_{\rm GCE}(k)\right]}{\partial \mu}}
is the GCE average number of the $k$th clusters.
The clusters pressure $p$ and particle number density $n$ can be found as 
\eq{
p &= (2\pi m)^{3/2}T^{5/2}\sum_{k \geq 1} k^{3/2} g(k) \exp\left(\frac{\mu k}{T}\right),\label{press}\\
n & =
 (2\pi m T)^{3/2}\sum_{k\ge 1} k^{5/2} g(k) \exp\left(\frac{\mu k}{T}\right).
}
Using Eqs.~(\ref{Pk}) and (\ref{press}) one can rewrite the pressure as  
\eq{p = \frac{nT}{\langle k\rangle}~,\label{p-cl}}
and the
scaled variance $\omega_{\rm gce}$ 
\eq{
\omega_{\rm gce} &= T\left[\frac{d p}{d n}\right]^{-1} = \frac{T}{n}\left(\frac{\partial n}{\partial \mu}\right)_T =\frac{\langle k^2\rangle }{\langle k\rangle}~, \label{omega-cl} 
}
where we defined $\langle k^l\rangle\equiv \sum_{k\ge 1}k^l P_k$.
Therefore, the first two moments of the  cluster probability distribution $P_k$ define both the system pressure (\ref{p-cl}) and scaled variance (\ref{omega-cl}). Due to the evident inequalities, $\langle k\rangle \ge 1$ and $\langle k^2\rangle \ge \langle k\rangle$, the results (\ref{p-cl}) and (\ref{omega-cl}) demonstrate that in the mixture of noninteracting $k$-th clusters ($k=1,2,\ldots$)   the system pressure becomes smaller and the scaled variance larger 
than the corresponding ideal gas values $p_{\rm id}=nT$  and $\omega_{\rm id}=1$
with no cluster formation, i.e.,
when $g(k=1)=1,~g(k>1)=0$. General expression for cumulants $\kappa_n[N]$ of any order $n$ can be obtained:
\eq{
\kappa_n[N]= \frac{\partial^n \ln \left[Z_{\rm GCE}\right]}{\partial \left(\mu\over T\right)^n} =\mean{k^n}\sum\limits_{k \geq 1}\mean{ N_k}.
}

The  model of noninteracting clusters discussed above can be considered as an approximation for the LJ fluid in the nucleation region. The attractive part of the LJ potential is responsible for the $k$th cluster formation. On the other hand, the particle number density is still sufficiently small to justify the absence of the repulsive interaction effects between clusters.

By definition, a cluster is a bound system of particles. 
There are several ways to define clusters in molecular dynamics simulations
\cite{Sator2003, Moretto2011}. 
In the following, we will use the Hill algorithm~\cite{doi:10.1063/1.1742067}.  
A pair of particles $i$ and $j$ is assumed to be bound if their rest frame energy is negative,
\eq{\label{rf-e}
(\tilde v_i-\tilde v_j)^2+\tilde V_{LJ}(|\tilde r_i-\tilde r_j|)<0.
}
A given particle is assumed to belong to a cluster if it is bound to at least one other particle in that cluster.
Finding clusters is thus equivalent to finding connected components in an undirected graph, whose vertices correspond to particles and where all bound pairs of particles~[i.e. the condition (\ref{rf-e}) is satisfied] are connected by edges.
We use depth-first search~(DFS) to find the connected components of the graph and thus identify all the clusters.

Utilizing the above procedure, one obtains the probability distribution $P_k$ in a Lennard-Jones fluid for given $\tilde n$ and $\tilde T$ from MD simulations.  
Examples of the extracted $P_k$ distributions 
for $\tilde T=0.76$ and $\tilde T=1.06$ are shown in Figs. \ref{fig:Pm1} (a) and (b) for $\tilde n=0.16$ and $\tilde n=0.35$, respectively. 
The results indicate that cluster formation becomes more significant when either temperature $\tilde T$ is decreased or particle number density $\tilde n$ is increased.

\begin{figure}[h!]
    \includegraphics[width=.49\textwidth]{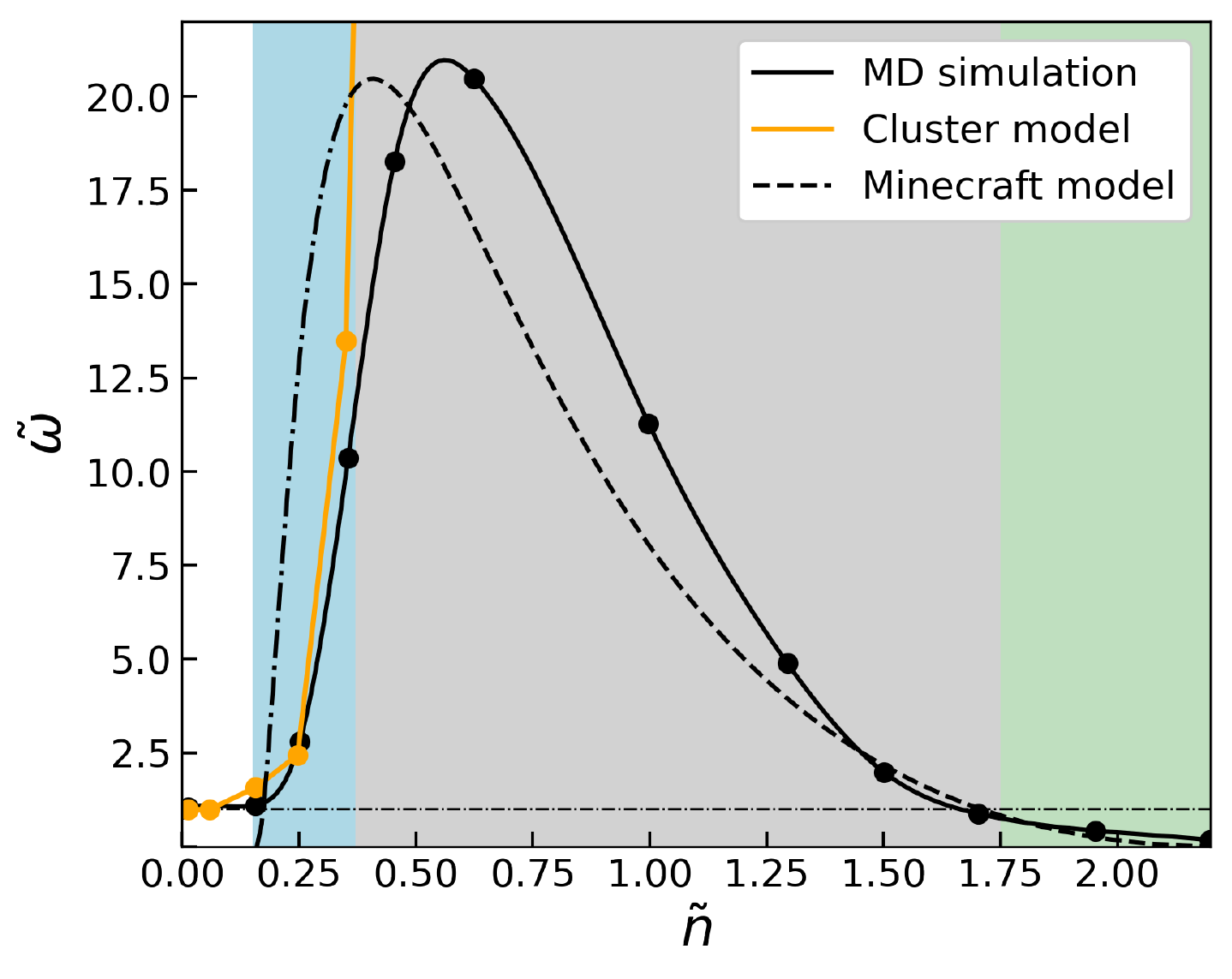}
    \caption{
    The points connected by the solid line correspond to the MD results for $N_0=400$ and $\alpha =0.2$ at $\tilde T=0.76$. The orange line demonstrates the cluster model results in the nucleation region  $0.16\le \tilde n\le 0.35$. The dashed line shows the results of the Minecraft model in the spinodal region $0.35\le  \tilde n\le 1.75$, and the dash-dotted line is its 
    extension to the nucleation region. 
    }
    \label{fig:w00}
\end{figure}

We then use the extracted $P_k$ distributions to evaluate $\mean{k}$ and $\mean{k^2}$ which we then plug into (\ref{omega-cl}) to estimate the GCE scaled variance in the cluster model.
These results are compared with $\tilde \omega$ calculated in a subvolume $V=\alpha V_0$ directly from MD simulations. 
The cluster model results for $\tilde T=0.76$ are shown in Fig.~\ref{fig:w00} by the orange line. These results agree qualitatively with direct MD simulations data (black line) in the range of densities $0.16 \lesssim \tilde n \lesssim 0.35$ corresponding to the nucleation region.
In particular, cluster formation explains the strong rise (approximately by a factor of 20) of the scaled variance with $\tilde n$ in the nucleation region.
The cluster model, however, fails to describe the peak in $\tilde{\omega}$ seen in MD simulations at higher densities, indicating its breakdown in the spinodal region.

\section{Fluctuations in the spinodal region
}\label{SR}

In Ref.~\cite{Poberezhnyuk:2020cen}, the GCE particle number fluctuations were calculated in the mixed phase region. 
It was assumed that both the liquid and gas phases are entirely inside the system volume $V_0$ that tends to infinity.  
In MD simulations here, we instead study fluctuations in a subvolume $V=\alpha V_0$, which corresponds to a different scenario. 
We thus develop new models to understand qualitative features of the behavior observed in MD simulations.

In the spinodal region, one assumes that the volume $V$
is partitioned into volumes $V_l = x V$ and $V_g = y V$ occupied by the liquid and gaseous phases, respectively ($0<x<1$, $y \equiv 1-x$). The corresponding particle number densities in the liquid and gaseous phases are  $\rho_l\equiv N_l/V_l$ and $\rho_g=N_g/V_g$.
The $r$th moment of the particle number distribution in the subvolume $V=\alpha V_0$ can then be presented as the following:
\eq{\label{mf-moment}
\mean{N^r}=\mean{(N_l+N_g)^r}=V^r \, \mean{(x \rho_l+y \rho_g)^r}~.
}
The fluctuating quantities are the densities $\rho_l$, $\rho_g$, and the volume fraction $x$, whereas the  volume $V$ is fixed.
Following Refs.~\cite{Vovchenko:2015xja,Satarov:2020loq} we assume that
the fluctuations of all these quantities are independent in the thermodynamic limit, i.e., $\mean{\rho_l^k \, \rho_g^m \, x^n} = \mean{\rho_l^k} \, \mean{\rho_g^m} \, \mean{x^n}$ for any non-negative integers $k$, $m$, and $n$.

The first moment ($r = 1$),
reduces via Eq.~(\ref{mf-moment}) to
\eq{\label{mean-B}
\mean{N}=x_0 V n_l + y_0 V n_g = V n~,
}
where $x_0=\mean{x}$ is the mean volume fraction occupied by the liquid phase, $y_0\equiv 1-x_0$, and $n_l=\mean{\rho_l}$ and  $n_g=\mean{\rho_g}$ are the mean densities in the liquid and gaseous phases, respectively.
The particle number density is equal to $n \equiv \mean{N}/V=N_0/V_0$. Equation~(\ref{mean-B}) defines $x_0$ in terms of the mean densities:
\eq{\label{x0}
x_0\equiv\mean{x}=\frac{n-n_g}{n_l-n_g}~.
}
At fixed temperature $T<T_c$ the mean densities of the liquid $n_l$ and gaseous $n_g$ phases are assumed to remain constant 
with respect to system's particle number density $n$ in the spinodal region in the thermodynamic limit. These quantities coincide with the corresponding values on the liquid (right) and gaseous (left) binodals.  

Using Eq.~\eqref{mf-moment} one finds the variance of particle number distribution (see Ref.~\cite{Poberezhnyuk:2020cen} for details):
\eq{
&{\rm Var}[N] \equiv   \mean{N^2}-\mean{N^2} \nonumber\\ &= {\rm Var}_x[N_l]\left(1+\frac{{\rm Var}[x]}{x_0^2}\right)+{\rm Var}_x[N_g]\left(1+\frac{{\rm Var}[x]}{y_0^2}\right) \nonumber \\
& \quad +V^2(n_l-n_g)^2 {\rm Var}[x]~. \label{k2}
}
Here 
${\rm Var}_x[N_{l,g}]$
is the variance of $N_{l,g}$ at fixed volume fraction $x$ and  
${\rm Var}[x]$
is the variance of the $x$ distribution.

Suppose that  there are several blobs of the liquid and gaseous phases, and all of them  are much smaller than the 
subvolume  $V$. This would correspond to a spatially homogeneous mixed phase. 
In this case, ${\rm Var}[x]$ is expressed in terms of cumulants of $V_l$ distribution as
${\rm Var}[x] \equiv V^{-2}{\rm Var}[V_l]$.
In the thermodynamic limit, $V\rightarrow\infty$, all
cumulants of extensive quantities are proportional to the system volume,  ${\rm Var}_x[N_{\rm{l,g}}]\sim V$
and ${\rm Var}_x[V_{l,g}] \sim V$.
Eq.~\eqref{k2} reduces to 
\eq{
{\rm Var}[N] & = {\rm Var}_x[N_l] + {\rm Var}_x[N_g]\nonumber \\& + V^2(n_l-n_g)^2 {\rm Var}[x]~, \label{k22}
}
where all terms are linear in $V$.
The result (\ref{k22}) coincides with that obtained for the GCE in Ref.~\cite{Poberezhnyuk:2020cen},
and it corresponds to the finite values of the scaled variance at $T<T_c$ in the thermodynamic limit.  

Note that the above derivation is based on the assumption of homogeneity.
This assumption is valid for pure phases.
In the mixed-phase region, however, this assumption may only be reasonable when applied to long-lived metastable phases. 
Such a configuration of the system, however, can not be viewed as an equilibrium configuration in a region of spinodal decomposition.
There, the sizes of the liquid and gaseous blobs are both of the order of the total volume $V_0$, and their volumes are comparable to the subvolume $V$. Thus, the whole picture is strongly heterogeneous (see Fig.~\ref{fig:mixed} (c) and Fig.~\ref{fig:blob}). As a consequence, ${\rm Var}[x]$ becomes volume independent, thus, the last term in Eq.~\eqref{k2} is quadratic in $V$ and makes the dominant contribution to fluctuations.
Leaving only this last term, one obtains:
\eq{
&{\rm Var}[N] = V^2(n_l-n_g)^2 {\rm Var}[x]~, \label{k22}
}
and
\eq{\begin{split}\label{k23}
\tilde{\omega}[N]&=\frac{{\rm Var}[N]}{(1-\alpha)\mean{N}}\\
&=\alpha(1-\alpha)N_0\frac{(n_l-n_g)^2}{n^2}{\rm Var}[x]~.
\end{split}}
This result indicates that $\tilde{\omega}[N]$ scales with $N_0$, i.e. the scaled variance diverges in the thermodynamic limit. 
We checked that for $N_0 \gg 400$
the substantial increase of $\tilde \omega$ is observed within MD simulations, however the scaling behaviour for fluctuations is out of the scope of the present paper. 
In the following, we present estimates for ${\rm Var}[x]$.
 
{\bf Small $\alpha$ limit.}
At $\alpha \ll 1$ one has $V_{\rm l}\gg V$ and $V_{\rm g}\gg V$. 
This means that
one can neglect the events when both phases are simultaneously present inside the subvolume $V$,
and the whole subvolume is entirely inside either the gaseous or liquid phase.
The probability distribution $P[x]$ thus reads
\eq{
P[x] = x_0 \, \delta(1-x) + y_0 \, \delta(x)~.\label{Px}
}
This means that
one can neglect the events when both phases are simultaneously present inside the subvolume $V$. 
From Eq.~(\ref{Px}) one finds 
\eq{\label{deltas-var}
{\rm Var}[x]=x_0 y_0.
}
The maximal value of ${\rm Var}[x]=0.25$ is reached at $x_0=\sqrt[3]{0.5}$.
Using Eqs.~\eqref{x0} and~\eqref{k22} one obtains:
\eq{\label{omega-gce}
{\rm Var}[N] = V^2 (n - n_g)(n_l - n)~. 
}
One sees that the scaled variance of particle number distribution is indeed divergent inside the mixed phase in the thermodynamic limit, scaling with the subvolume $\tilde \omega\sim V$.

\begin{figure}
    \includegraphics[width=.40\textwidth]{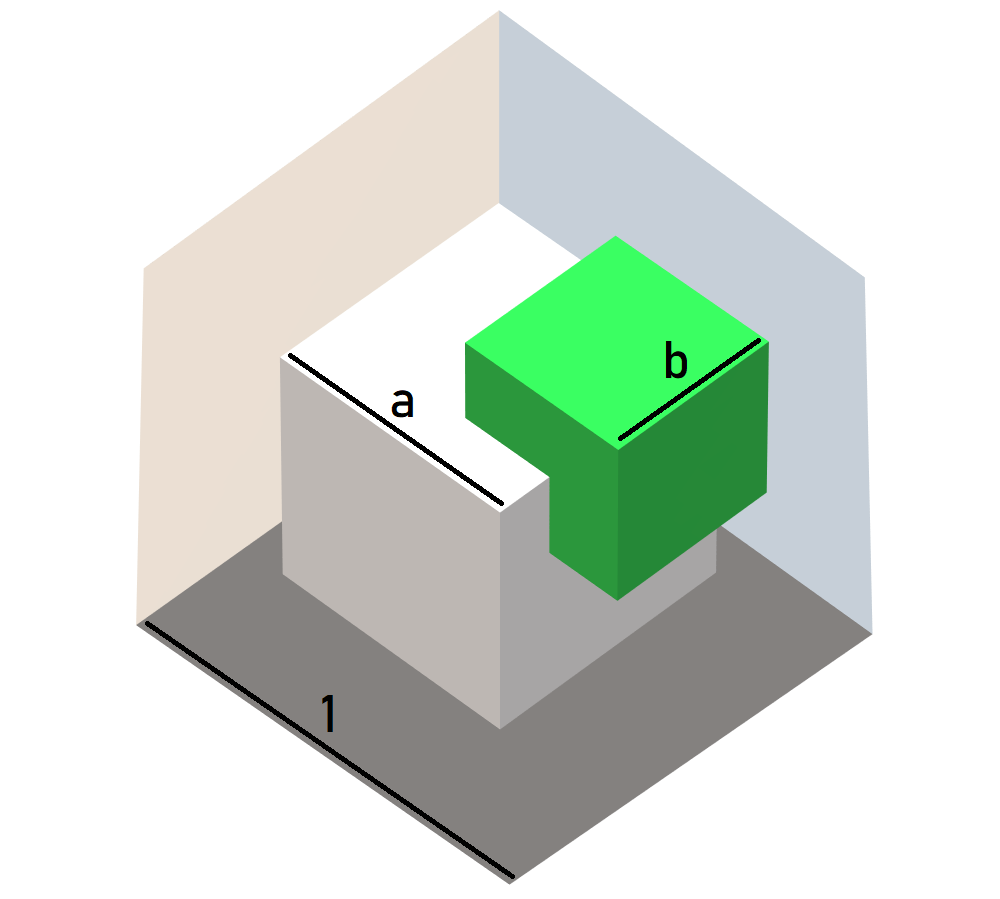}
    \caption{The illustration of the Minecraft toy model of an equilibrium system in the unstable region of the mixed phase. The subsystem is shown by the grey cube in the center while the green cube represents the liquid.
    The remaining space of the system is occupied by gas.}
    \label{cube}
\end{figure}

{\bf Minecraft model.}\footnote{This name is inspired by the popular 3D video game.}
Now let us calculate ${\rm Var}[x]$ when the sizes of the volume, subvolume, and blobs are all comparable.
For that we consider a simple "geometric" toy model of the cubic system with unit volume which contains both liquid and gaseous phases (see Fig.~\ref{cube}). The cubic subvolume $V=\alpha$ is located in the center of the system with coordinates $(a_{\rm x},a_{\rm y},a_{\rm z})=(0,0,0)$. The edge length of the subvolume is $a=\sqrt[3]{\alpha}$. All liquid is condensed into a single blob which freely moves within the system.  Here we neglect the effects of a geometric form  and assume that this blob has a shape of a perfect cube. The volume of the cube of liquid is $V_l=x_0$. Correspondingly its edge length is $b=\sqrt[3]{x_0}$. 
The system has periodic boundary conditions, therefore, the coordinates $(b_x,b_y,b_z)$ of the center of the cube of liquid are limited by $-\frac{1}{2}<b_{\rm x},b_{\rm y},b_{\rm z}<\frac{1}{2}$.
The fraction $x$ of the subvolume occupied by the liquid phase is the overlap volume between the cubic subvolume and the cubic liquid divided by the subvolume $V=\alpha$.

The system has three degrees of freedom -- the coordinates of the liquid cube $(b_x,b_y,b_z)$.
Since the cube center is uniformly distributed over $-\frac{1}{2}<b_{\rm x},b_{\rm y},b_{\rm z}<\frac{1}{2}$, the three coordinates are independent.
The fraction $x$ as a function of these three coordinates and can be written as
\eq{
x = \frac{f(b_{\rm x})f(b_{\rm y})f(b_{\rm z})}{\alpha}.
}
Here
$f(b_{\rm i})$ is the overlap of liquid blob with subvolume along the coordinate $i$ as a function of $b_{\rm i}$.

\begin{figure}
    \includegraphics[width=.39\textwidth]{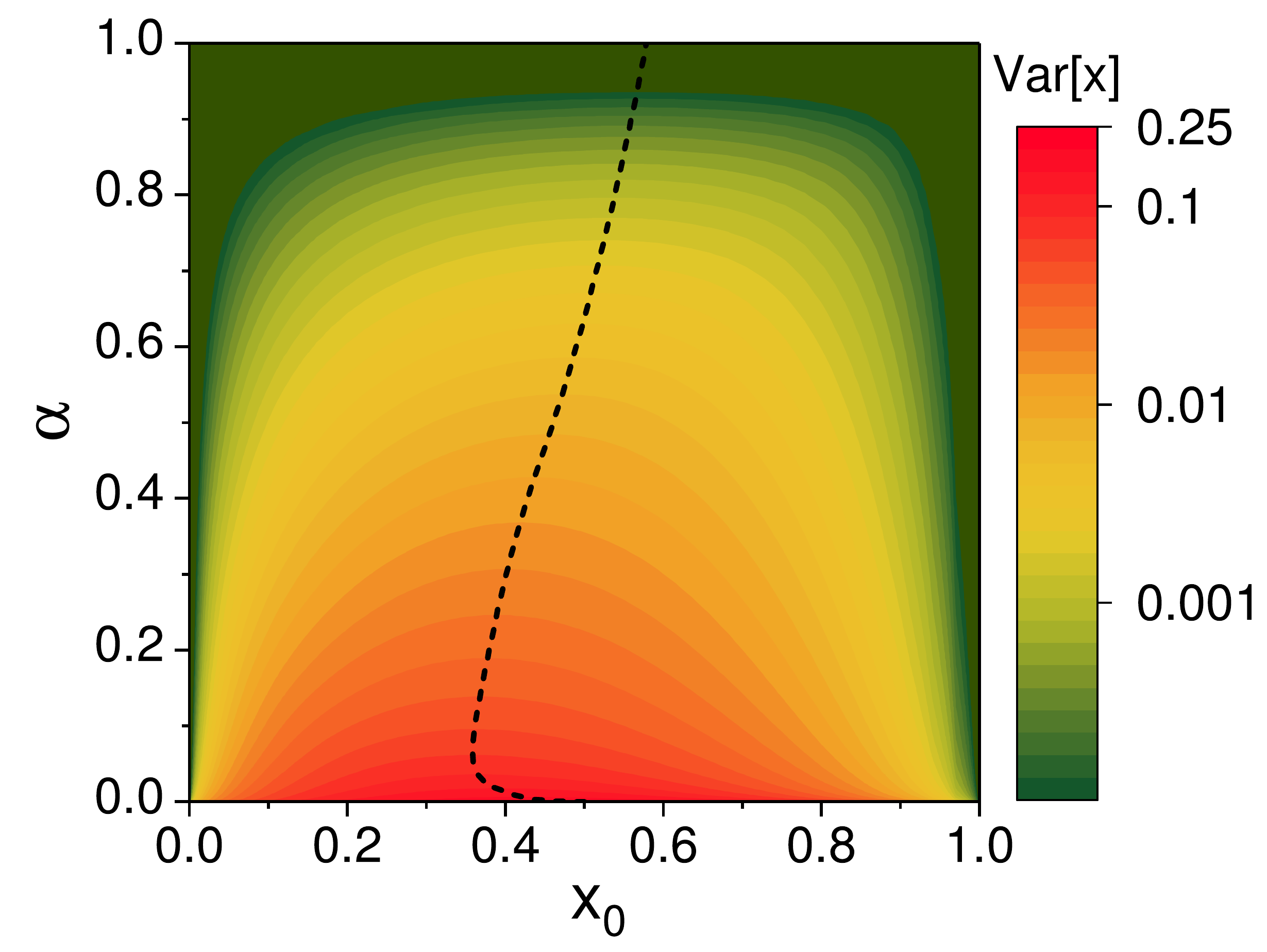}
    \caption{The variance of the volume fraction occupied by the liquid phase, ${\rm Var}[x]$, as a function of $\mean{x}\equiv x_0$ and $\alpha$ calculated in the Minecraft model~[Eq.~(\ref{cube-var})]. 
    The dashed line corresponds to the maximum value of ${\rm Var}[x]$ at fixed $\alpha$.
    }
    \label{fig-varx}
\end{figure}

The mean value $\mean{x}$ can be found as 
\eq{\label{mean-x}
\mean{x}=\frac{1}{\alpha v}\left(\int_{-1/2}^{1/2} f(b_{\rm i}) {\rm d}b_{\rm i}\right)^3=x_0
}
where $v=1$ is the volume of the system.
Similarly, one can calculate the variance of $x$:
\eq{\label{cube-var}
&{\rm Var}[x]=-b^6 \\\nonumber
&+\left[\frac{b^2 (3a-b)+\Theta_{a+b-1}(a+b-1)^3+\Theta_{b-a}(b-a)^3}{3 a^2}\right]^3
}
where $\Theta_{...}\equiv\Theta[...]$ is the 
step function and, as before, $a=\sqrt[3]{\alpha}$ and $b=\sqrt[3]{x_0}$.
One sees that Eq.~(\ref{cube-var}) reduces to Eq.~(\ref{deltas-var}) when $\alpha\rightarrow 0$.
In other limiting cases ${\rm Var}[x]\rightarrow 0$ when $\alpha\rightarrow 1$ or $x_0\rightarrow 1$ or $x_0\rightarrow 0$. ${\rm Var}[x]$ as a function of $x_0$ and $\alpha$ is shown in Fig.~\ref{fig-varx}.

The scaled variance $\tilde{\omega}[N]$ 
given by Eq.~(\ref{k23}), with ${\rm Var}[x]$ estimated using the Minecraft model, Eq.~(\ref{cube-var}), is shown in Fig.~\ref{fig:w00} in spinodal and nucleation  regions by  dashed and dotted lines, respectively. 

\section{Summary}\label{res}
We studied particle number fluctuations inside the mixed phase of a liquid-gas phase transition by utilizing molecular dynamics simulations of the Lennard-Jones fluid.
The simulations were performed for $N_0 = 400$ particles in a cubic box with periodic boundary conditions.
The fluctuations are studied inside a cubic subvolume $V= 0.2\,V_0$ located in the geometrical center of the system.

First, we briefly explore the supercritical temperature, where one observes the approximate Gaussian shape of the $P(N)$ distribution. 
The scaled variance $\tilde \omega$ characterizes the width of the $P(N)$ distribution.
It first increases with density
from $\tilde \omega\approx 1$ at small $\tilde n$ to its maximum above unity around the critical density $\tilde n=1$,
and then it decreases with $\tilde n$ to small values $\tilde \omega <1$. This is illustrated in Fig.~\ref{fig:w0} (b).

The situation differs in the mixed phase, $\tilde T < 1$.
The structure of the $P(N)$ distribution is significantly more intricate.
For $\tilde n\approx 1$, the distribution is bi-modal, as 
shown in  
Fig. \ref{fig:p1}. 
The corresponding variance of particle number is much more significant compared to pure phases~[Fig.~\ref{fig:w0}~(a)]. 

To understand the qualitative features of the observed behavior, we formulate two phenomenological toy models.

The first model describes the system as non-interacting multi-component gas of $k$-particle clusters,
taking the cluster probability distribution $P_k$ directly from the MD simulations as input.
The model describes semi-quantitatively the rapid increase of $\tilde \omega$ with density in the nucleation region of the mixed phase, 
i.e., the region between the gaseous binodal and spinodal~[Fig.~\ref{fig:w00}].

The second model -- the Minecraft model -- is formulated for the spinodal region of the mixed phase.
The particles are separated
into two phases, namely, the liquid blob with volume $V_{\rm{l}}$ surrounded by gas.
The size of the blob $V_{\rm{l}}$ can be expressed through the total density of the system $\tilde n$ and densities on the binodals.
The Minecraft model considers the geometrical effects that become important when the volumes $V_{\rm{l,g}}$ and $V$ are of comparable size. 
With this consideration, the model indicates that $\tilde \omega \sim N_0\rightarrow\infty$, thus the variance is divergent in the thermodynamic limit inside the spinodal region.


The present work is motivated by the study of event-by-event fluctuations in nucleus-nucleus collisions to probe the phase structure of QCD.
Our MD simulations inside the mixed phase were performed for $400$ particles, while the fluctuations were studied in the subvolume $V=0.2\,V_0$. 
These two parameters correspond to typical total numbers of nucleons and the percentage of accepted final particles in heavy-ion collisions.
The results indicate that large fluctuations of particle number in coordinate space can be interpreted as a signal of the spinodal region of the first-order phase transition.
However, there are significant differences between our calculations and heavy-ion collisions.
One difference is that in heavy-ion collisions, particles are not detected during the equilibrium phase of the collision but only after they fly away to the detector.
Another difference is that particle momenta, not the coordinates, are measured in the experiment.
We plan to address these issues by performing MD simulations of expanding systems. 

Our simulations provide a first microscopic model test of the subensemble acceptance method~(SAM)~\cite{Vovchenko:2020tsr,Vovchenko:2020gne} in the mixed phase region of a first-order phase transition. 
The SAM is a method for correcting the baryon number cumulants in heavy-ion collisions, which is model-independent in the thermodynamic limit, and it was previously tested in the crossover region at supercritical temperatures in Ref.~\cite{Kuznietsov2022}.
Our simulations reveal that the SAM remains accurate in metastable regions of the phase diagram but breaks down in the spinodal decomposition region. The reason is that the finite-size effects remain sizable even in large systems in this region of the phase diagram. The treatment of the canonical effects in the spinodal region is thus more complex. It will require appropriate generalizations of the SAM, such as including macroscopic geometrical effects encompassed by the Minecraft model introduced here.

Another future avenue is generalizing the presented analysis to higher-order moments of particle number distributions, such as skewness and kurtosis.

\section*{Acknowledgements}
The authors are thankful to Jeroen van den Brink, Volker Koch, Flavio Nogueira, Scott Pratt and Jan Steinheimer for fruitful comments and discussions. O.S. acknowledges the scholarship grant from the GET$\_$INvolved Programme of FAIR/GSI and support by the Department of Energy Office of Science through grant no. DE-FG02-03ER41259. This work is supported by the National Academy of Sciences of Ukraine, Grant No.  0122U200259.
 M.I.G. and R.V.P. acknowledge the support from the Alexander von Humboldt Foundation. This work was supported by a grant from the Simons Foundation (Grant Number 1039151). H.St. appreciates the Judah M. Eisenberg Professur Laureatus of the Walter Greiner Gesellschaft/F\"orderverein f\"ur physikalische Grundlagenforschung Frankfurt, and the Fachbereich Physik at Goethe Universit\"at.
 
\appendix

\section{Cluster partition function in the CE}
\label{A}

For a system of non-interacting multi-component gas of $k$th particle clusters, the canonical ensemble (CE) partition function reads

\eq{
Z_{\rm CE} (V,T,N) &= 
\prod_{k = 1}^N  \sum_{N_k\ge 0} \frac{\left(Vg(k)\right)^{N_k}}{N_k!}(2\pi k m T)^{3N_k/2} \nonumber \\ 
& \times \delta \left[N -\sum_{k=1}^{N} kN_k \right]~.
\label{sumZCE}
}

Applying the integral form of the Kronecker symbol,
\eq{\delta \left[N- \sum_{k=1}^N kN_k \right] 
= \int\limits_0^{2\pi} \frac{d \varphi}{2\pi}
\exp \left[i\varphi\left(N- \sum_{k=1}^N kN_k\right)  \right] ~,\label{ds}
}

to Eq.~(\ref{sumZCE}), one obtains
\eq{
Z_{\rm CE} (V,T,N)= 
\int \limits^{2\pi}_0 \frac{d\varphi}{2\pi} ~ e^{-i\varphi N}  \exp \left[ \sum_{k \geq 1} r(k) e^{i\varphi k}\right].
\label{ZCE0}
}

Here $r(k) \equiv V g(k) (2\pi k m T)^{3/2}$.

Using the Maclaurin expansion, one has
\eq{
\exp \left[ \sum_{k \geq 1} r(k) e^{i\varphi k}\right] = \sum_{l = 0}^{\infty} \frac{B_l(r(1),\dots,l!r(l))}{l!} e^{i\varphi l},\label{fab}}
where $B_l$ are Bell polinomials \cite{Bell1927}.

Substituting \eqref{fab} into \eqref{ZCE0} gives 
\eq{
Z_{\rm CE}(V,T,N) = 
\frac{B_N(r(1),\dots,N!r(N))}{N!}~.\label{Zce}
}

From the above equations one finds the GCE partition function

\eq{Z_{\rm GCE} = \sum_{N=0}^{\infty} Z_{\rm CE} \exp\left(\frac{\mu N }{T}\right) = \prod_{k = 1}^N\exp\left(r(k)e^{\mu k/T}\right)~,\label{Zgce1}}
which coincides with Eq.~(\ref{Zgce0}).
$Z_{\rm CE}$ can be expressed in terms of $Z_{\rm GCE}$ through the Mellin transformation 

\eq{Z_{CE} = \int\limits^{c + i\infty}_{c-i\infty} Z_{GCE}e^{-\mu N/T} d\mu ~.\label{Me}}

The integral \eqref{Me} can be evaluated in the large $N$ limit using the steepest descent method \cite{SSm}.
Therefore,

\eq{
Z_{\rm CE} (V,T,N)~ &\approx ~\sqrt{\frac{2\pi T^2}{\sum_{k = 1}^N k^2r(k)e^{\mu_0 k/T}}} \nonumber \\
& \times~\exp\left(\sum_{k = 1}^N r(k) e^{\mu_0 k/T} - \frac{\mu_0}{T} N\right)~, \label{ZCE-1}
}
where $\mu_0(T,N)$ can be found from the saddle point  equation
\eq{\sum_{k=1}^N k r(k)e^{\mu_0 k/T} - N = 0~.}

Equation~(\ref{ZCE-1}) indicated that the $j$-th moment of $k$th cluster distribution in the large $N$ limit reads

\eq{\left<k^j\right>_{\rm CE} ~ = ~
\left<k^j\right>_{\rm GCE} ~+ ~ O(N^{-1})~.}

This result shows that all moments $j=1,2,\ldots$ of the $k$-th cluster distribution $(k=1,\ldots,N)$ are the same in the CE and GCE in the thermodynamic limit $N\rightarrow \infty$. 

If $N$ is large, this justifies the use of $P_k$ probabilities from MD simulations as input into the calculations of fluctuations in the GCE using formulas (\ref{p-cl}) and (\ref{omega-cl}).
\bibliography{references.bib}
\end{document}